\documentclass{svproc}
\usepackage{url}

\usepackage[hidelinks]{hyperref}

\usepackage{amsmath}

\usepackage[linesnumbered,ruled]{algorithm2e}

\SetArgSty{textnormal}

\SetKwInput{KwInput}{Input}                
\SetKwInput{KwOutput}{Output}              

\usepackage[style=nature]{biblatex}
\bibliography{references.bib}

\begin{document}
\mainmatter              
\title{On social simulation in 4D relativistic spacetime}
\titlerunning{On social simulation in 4D relativistic spacetime}  
%
\author{Lai Kwun Hang}
\authorrunning{Lai Kwun Hang} 
%
%
\institute{Centre for Science and Technology Studies (CWTS), Leiden University,\\
\email{k.h.lai@cwts.leidenuniv.nl}
}

\maketitle              

\begin{abstract}
Agent-based modeling and simulation allow us to study social phenomena in hypothetical scenarios.
If we stretch our imagination, one of the interesting scenarios would be our interstellar future.
To model an interstellar society, we need to consider relativistic physics,
which is not straightforward to implement in existing agent-based simulation frameworks.
In this paper, we present the mathematics and algorithmic details needed for simulating agent-based models in 4D relativistic spacetime.
These algorithms form the basis of our open-source computational framework, ``Relativitization''~\cite{relativitization}.
\keywords{special relativity, interstellar society, agent-based simulation, computational framework}
\end{abstract}

\section{Introduction}
The scientific and technological advancement in the last century greatly increases our understanding of the universe.
Nowadays, we are able to build giant telescopes and observe astronomical objects billions of light years away.
Apart from deepening our scientific understanding, our astronomical knowledge also stimulates our imagination
of interstellar civilizations.
A lot of great science fiction has been written, and scientists have proposed ideas like 
the Fermi paradox~\cite{gray2015fermi}, the Dyson sphere~\cite{wright2020dyson} and the Kardashev scale~\cite{gray2020extended}.
While many of these ideas are physically plausible, 
it would be interesting to discuss these ideas from a social science perspective.
Due to the highly hypothetical nature of the problem,
we suggest that agent-based model (ABM) enables formal academic discussion on interstellar society.

An ABM is a simulation model bridging microscopic behaviours of agents and macroscopic observations.
Depending on the context of the model, an agent can be an individual, an organization, or even a country.
First, assumption about the social behaviours of the agents are made, then the agents are placed and evolved
in a computational environment.
For a research problem for which it is not viable to perform data collection and analysis,
ABM can still be used for theoretical exposition~\cite{edmonds2015simulating}.

To model agents in interstellar space, suppose we only consider a scale with normal stellar objects
so that we can ignore the effects of general relativity, such as universe expansion and black holes,
we still have to consider the effect of special relativity.
In the context of ABM, where the simulation is computed under a set of inertial frames that are at rest to each other,
we can simplify the theory of relativity into two core
phenomena: speed of light as the upper bound of the speed of information travel, and
time dilation relative to any stationary observer in the inertial frames.
This also implies that we have to take care of four dimensions: three space dimensions, plus one time dimension.

Typically, an ABM is constructed using an ABM framework to facilitate model development and communication.
There are a lot of existing ABM frameworks, to name a few, NetLogo~\cite{netlogo}, mesa~\cite{python-mesa-2020},
and Agents.jl~\cite{Agents2021}, see~\cite{pal2020review} for a detailed review.
While it is possible to build a 4D relativistic model in some existing ABM frameworks, 
those frameworks do not have native support for the necessary 4D data structures,
and it can be error-prone to enforce relativistic effects via custom implementations of data structures and algorithms.
Therefore, we have developed a simulation framework we call ``Relativitization''~\cite{relativitization},
to help social scientists to build an ABM in relativistic spacetime.
In this paper, the mathematics and the algorithms underlying the framework will be presented.

\section{Definitions}

In Relativitization, an agent is called a ``player''.
Players live in a universe.
Ideally, computation should be done in every local frame following all players, 
and the computation results can be synchronized by Lorentz transformations.
However, this will make the framework and the model substantially more complex.
Therefore, all computations are done according to some inertial frames that are at rest to each other.
The spatial coordinates of a player are represented by floating-point numbers $x$, $y$ and $z$
and the time coordinate of a player is represented by a floating-point number $t$.
To simplify computation and visualization, the universe is partitioned into unit cubes.
A player with floating-point coordinates $(t, x, y, z)$ is located at the cube 
with integer coordinates $T = \lfloor t \rfloor$, $X = \lfloor x \rfloor$, $Y = \lfloor y \rfloor$, $Z = \lfloor z \rfloor$,
note that the computations of a simulation are done at unit time steps and we can actually assume $T = t$.
Denote the speed of light as $c$.
In vector notation, define $\textbf{s} = (t, \overrightarrow{u}) = (t, x, y, z)$, and 
$\textbf{S} = (T, \overrightarrow{U}) = (T, X, Y, Z)$.

\subsection{Interval and time delay}

The spacetime interval between coordinates $\textbf{s}_i$ and $\textbf{s}_j$ is
\begin{equation}
    \|\textbf{s}_i - \textbf{s}_j\| = c^2 (t_i - t_j)^2 - (x_i - x_j)^2 - (y_i - y_j)^2 - (z_i - z_j)^2.
\end{equation}

If $\|\textbf{s}_i - \textbf{s}_j\| < 0$, it is called a spacelike interval, and events that happen at the two coordinates
are not causally connected because no information can travel faster than the speed of light $c$.

It is often needed to compute intervals in integer coordinates.
We define the spatial distance between $\overrightarrow{U_i}$
and $\overrightarrow{U_j}$ as the maximum distance between all points in the cubes at 
$\overrightarrow{U_i}$ and $\overrightarrow{U_j}$
\begin{equation} \label{eq:delay}
    |\overrightarrow{U_i} - \overrightarrow{U_j}| = (X_i - X_j + 1)^2 + (Y_i - Y_j + 1)^2 + (Z_i - Z_j + 1)^2.
\end{equation}

Suppose there is a signal sent from $\overrightarrow{U_i}$ to $\overrightarrow{U_j}$.
To ensure that the information travels slower than the speed of light, 
the integer time delay $\tau(\overrightarrow{U_i}, \overrightarrow{U_j})$ is computed as
\begin{equation} 
    \tau(\overrightarrow{U_i}, \overrightarrow{U_j}) = \left \lceil \frac{|\overrightarrow{U_i} - \overrightarrow{U_j}|}{c} \right \rceil.
\end{equation}

\subsection{Group id}

From Eq.~\ref{eq:delay}, even if $\overrightarrow{U_i} = \overrightarrow{U_j}$, the time delay is non-zero.
To implement zero time delay for players that are really close to each other, 
we divide a unit cube into several sub-cubes with edge length $d_e$, 
and information travel within the same sub-cubes is instantaneous.

To improve the computational speed when checking whether two players belong to the same sub-cube,
we assign a ``group id'' to each sub-cube in a unit cube.
A unit cube has $n_e^3$ sub-cubes, where $n_e = \left \lceil \frac{1}{d_e} \right \rceil$.
For a player at $\overrightarrow{u}$, it belongs to the $(n_x, n_y, n_z)$ sub-cubes,
where $n_x = \left \lfloor \frac{x - X} {d_e} \right \rfloor$,
$n_y = \left \lfloor \frac{y - Y} {d_e} \right \rfloor$,
and $n_z = \left \lfloor \frac{z - Z} {d_e} \right \rfloor$.
The group id $g(\overrightarrow{u}, \overrightarrow{U})$ of the player
can be computed as
\begin{equation} \label{eq:group}
    g(\overrightarrow{u}, \overrightarrow{U}) = n_x n_e^2 + n_y n_e + n_z.
\end{equation}

If two players have the same integer coordinates and the same group id,
then we say the players belong the same group and the time delays between the players are zero.

\subsection{Player data}

A player is characterized by a set of data:
\begin{itemize}
  \item player id $i$,
  \item integer coordinates $(T_i, X_i, Y_i, Z_i)$,
  \item a historical record of integer coordinates $H_i = \{(T_i', X_i', Y_i', Z_i') \mid T_i' < T_i \}$,
  \item floating-point coordinates $(t_i, x_i, y_i, z_i)$,
  \item time dilation counter variables $\mu_i$, a floating point variable,
        and $\nu_i$, a boolean variable,
        to keep track of time dilation (see Sec.~\ref{ssec:mechanism} and Sec.~\ref{ssec:mechanisms}),
        $\mu_i = 0$ initially for all players,
  \item group id $g_i$,
  \item floating-point velocities $\overrightarrow{v_i} = (v_{ix}, v_{iy}, v_{iz})$,
  \item other data $D_i$ relevant to the model.
\end{itemize}

\subsection{Command}

In other frameworks, interactions in ABMs are often presented as one player asking another player to do something.
Because the speed of information travel is bounded by $c$,
a player cannot simply ask other players to do something immediately.
Instead, interactions are mediated by commands.
Whenever player $i$ wants to interact with player $j$, 
player $i$ sends a command to $j$. 

A command is characterized by:
\begin{itemize}
  \item $i_{\textrm{to}}$, the id of the player to receive this command,
  \item $i_{\textrm{from}}$ the id of the target player who sent this command,
  \item $\textbf{S}_{\textrm{from}}$ the integer coordinates when the player sent this command,
  \item $f_{\text{target}}$ a function to modify data of the target player when this is received.
\end{itemize}

Commands travel at the speed of light $c$.
The amount of time needed for a command to reach the target, 
measured in the inertial frames we used in the simulation,
depends on the trajectory of the target player $i_{\textrm{to}}$ and the sender coordinates $\textbf{S}_{\textrm{from}}$,

\subsection{Universe data}

Universe is an overarching structure which aggregates all necessary data and functionalities.
An universe has:
\begin{itemize}
  \item a current universe time $T_{\textrm{current}}$,
  \item a 4-dimension array of maps from player id to lists of player data $M_{TXYZ}$,
        so that the data of a player residing at $(T, X, Y, Z)$ is stored in the associated list, 
        the ``afterimages'' of players are also stored in the corresponding list (Sec~\ref{ssec:move}),
  \item a map $M_{\textrm{command}}$ from player id to lists of commands,
        such that a command in the list will be executed when the player receive the command,
  \item other universe global data $D_G$ relevant to the model,
\end{itemize}

\subsection{Mechanism} \label{ssec:mechanism}

Given an instance of a universe, 
the dynamics of players are based on predefined rules and the state of the universe observed by the players.
In our framework, we call the rules mechanisms.
A mechanism takes the state of the universe observed by a player,
modifies the state of a player,
and generates a list of commands to send to other players.

To ease the model development to account for the time dilation effect, 
we further divide mechanisms into two categories: regular mechanisms and dilated mechanisms. 
A regular mechanism is executed once per turn, 
while a dilated mechanism is executed once per multiple turns, 
adjusted for the time dilation of the player measured in the inertial frames we used in the simulation.

\section{Simulation step} \label{sec:simulation}

The following are needed to define a model:
\begin{itemize}
  \item the data structure of other player data $D_i$,
  \item the data structure of other universe global data $D_G$,
  \item a set of available commands,
  \item a function to initialize the universe data,
  \item a function to update the universe global data,
  \item a set of regular and dilated mechanisms,
\end{itemize}

Along with the universe data,
it is useful to define a map $M_{\textrm{current}}$ from a player id to the current player data, 
i.e., $T_i = T_{\textrm{current}}$,
as an internal object of the simulation.
The modifications of player data are first performed on $M_\textrm{current}$, 
and then synchronized back to the universe data at appropriate timing.

Suppose we have initialized an universe model and $M_\textrm{current}$, a complete step in a simulation involves:
\begin{enumerate}
  \item update the global data (Sec.~\ref{ssec:global}),
  \item compute time dilation effects for all players (Sec.~\ref{ssec:dilation}),
  \item process mechanisms for each player (Sec.~\ref{ssec:mechanisms}),
  \item process the command map (Sec.~\ref{ssec:command}),
  \item move players, add afterimages, and update time (Sec.~\ref{ssec:move}).
\end{enumerate}

The simulation can be ran for a fixed amount of steps, or stop when a stopping condition is met.

\subsection{Update global data} \label{ssec:global}

A model may rely on a mutable global data $D_G$ to implement the dynamics.
If the model depends on some player data to update $D_G$, 
and the effect is observable by players,
we need to ensure that no information is transferred faster than the speed of light via
the global data update.

For example, if the global data is modified if ``all'' player data satisfy a condition,
we have to be careful about what we mean by ``all'' here.
In the universe, the maximum time delay equals $\tau_{\textrm{max}} = \tau((0, 0, 0), (max(X), max(Y), max(Z)))$.
To fulfill the speed of light constraint, the update function has to check whether all player data in
$M_{TXYZ}$, where $T_{\textrm{current}} - \tau_{\textrm{max}} \leq T \leq T_{\textrm{current}}$,
$0 \leq X \leq max(X)$, $0 \leq Y \leq max(Y)$, and $0 \leq Z \leq max(Z)$,
satisfy that condition.

\subsection{Compute time dilation} \label{ssec:dilation}

Relative to a stationary observer $j$ in an inertial frame,
special relativity predicts that a moving observer $i$ experiences a time dilation effect:

\begin{equation} \label{eq:gamma}
  \gamma_i = \frac{1}{\sqrt{1 - \frac{v_i^2}{c^2}}},
\end{equation}

\begin{equation}
  \Delta t_i = \frac{\Delta t_j}{\gamma_i},
\end{equation}
where $\gamma_i$ is called the Lorentz factor.

To account for the time-dilation effect, 
the time dilation counter variables $\mu_i$ and $\nu_i$ are updated by Algorithm~\ref{alg:dilation}
every turn for every player. 
$\nu_{i}$ will then affect the mechanism processing in Sec.~\ref{ssec:mechanisms}.

\begin{algorithm}
\KwInput{$M_\textrm{current}$, map from player id to current player data}

\ForEach{player $i$ in $M_\textrm{current}$}{
  $\mu_i \gets \mu_i + \sqrt{1 - \frac{v_i^2}{c^2}}$\;
  \If{$\mu_i \geq 1$} {
    $\mu_i \gets \mu_i - 1$\;
    $\nu_i \gets \textbf{true}$\;
  }
  \Else {
    $\nu_i \gets \textbf{false}$\;
  }
}

\caption{Update time dilation counter.}
\label{alg:dilation}
\end{algorithm}

\subsection{Process mechanisms} \label{ssec:mechanisms}

Before processing any mechanism for a player, 
we need to compute the state of the universe viewed by the player.
At an instance in our discretized relativistic universe,
player $i$ sees other players located at the unit cubes closest
to the surface of the past light cone of player $i$,
while the entire cubes are still within the past light cone.
The computation consists of two steps:
(1) Algorithm~\ref{alg:viewAtCube} computes the view centered at a specific cube, 
ignoring the zero time delay when players are within the same group,
(2) Algorithm~\ref{alg:viewAtGroup} computes the view for players in a group.
Assuming each line of these algorithms takes $O(1)$ and iterating over all $(X, Y, Z)$,
the time complexity $O(mn)$ from Algorithm~\ref{alg:viewAtCube} dominates,
where $m=X_{\textrm{max}} Y_{\textrm{max}} Z_{\textrm{max}}$ is the spatial size of the universe,
and $n$ is the number of player.

\begin{algorithm}
\KwInput{\\
  \Indp
  $T_i, X_i, Y_i, Z_i$ position of the viewing location\\
  $M_{TXYZ}$ 4D array of maps from player id to lists of player data
}
\KwOutput{\\
  \Indp
  $M$ map from player id to player data\\
  $\Lambda_{XYZ}$ 3D array of maps from group id to lists of player id
}

Initialize empty $M$ and $\Lambda_{XYZ}$\;
\ForEach{$X_j, Y_j, Z_j$}{
  $T_j \gets T_i - \tau(\overrightarrow{U_i}, (X_j, Y_j, Z_j))$\;
  \ForEach{player data in $M_{T_j X_j Y_j Z_j}[k]$}{
    \If{$M$ has key $k$}{
      \If{$T$ of $M[k]$ < $T$ of the new player data}{
        Replace $M[k]$ by this new player data\;
      }
    }
    \Else{
      Store data of player $k$ to $M[k]$\;
    }
  }
}
Associate the player id from $M$ to the corresponding list in $\Lambda_{XYZ}$ by spatial coordinates and group id\;
\Return $(M, \Lambda_{XYZ})$

\caption{Compute the view of the universe at a cube, ignore the zero time delay when player are in the same group}
\label{alg:viewAtCube}
\end{algorithm}

\begin{algorithm}
\KwInput{\\
  \Indp
  $g_i$ group id\\
  $T_j, X_j, Y_j, Z_j$ position of the viewing location\\
  $M$ map from player id to player data\\
  $\Lambda_{XYZ}$ 3D array of maps from group id to lists of player id\\
  $M_{TXYZ}$ 4D array of maps from player id to lists of player data
}
\KwOutput{\\
  \Indp
  $M'$ map from player id to player data\\
  $\Lambda'_{XYZ}$ 3D array of maps from group id to lists of player id
}

$M' \gets M$\;
$\Lambda'_{XYZ} \gets \Lambda_{XYZ}$\;
\ForEach{player data in $M_{T_j X_j Y_j Z_j}[k]$ where $g(\overrightarrow{u_k}, \overrightarrow{U_k}) = g_i$}{
  \If{$T$ of $M[k]$ < $T$ of the new player data}{
    Replace $M'[k]$ by this new player data\;
    Update the corresponding position of player $k$ in $\Lambda'_{XYZ}$\;
  }
}
\Return $(M', \Lambda'_{XYZ})$

\caption{Compute the view of the universe for players in a group.}
\label{alg:viewAtGroup}
\end{algorithm}

The view of the universe of a player is used by mechanisms to update the player data and generate commands to send.
Regular mechanisms update the data of the player each turn,
while dilated mechanisms update the player if the time dilation counter variable $\nu_i$ is \textbf{true}
to account for the time dilation effect.
The generated commands are executed immediately if the target player is within the same group of the sender,
otherwise the commands are stored in $M_{\textrm{command}}$.
Algorithm~\ref{alg:mechanisms} shows the overall iterative process.

\begin{algorithm}
\KwInput{\\
  \Indp
  $M_\textrm{current}$ map from player id to current player data\\
  Universe data
}

\ForEach{$(X_j, Y_j, Z_j)$}{
  Compute the view of the universe at this cube by algorithm~\ref{alg:viewAtCube}\;
  \ForEach{group in this cube}{
    Compute the view of the universe at this group by algorithm~\ref{alg:viewAtGroup}\;
    \ForEach{data of player $k$ in this group}{
      Update $M_\textrm{current}[k]$ by all regular mechanisms\;
      \If{$\nu_k$ is \textbf{true}}{
        Update $M_\textrm{current}[k]$ by all dilated mechanisms\;
      }
    }
    \ForEach{generated command where target player $l$ is in this group}{
      Update $M_\textrm{current}[l]$ by $f_{\text{target}}$ of the command\;
    }
  }
}
Add the rest of commands to $M_{\textrm{command}}$ by the target player id of the commands\;

\caption{Update all players by mechanisms.}
\label{alg:mechanisms}
\end{algorithm}

\subsection{Process command map} \label{ssec:command}

The command map $M_{\textrm{command}}$ is a map from player id to lists of commands that is being sent to that player.
At each turn, the distance between the target player and the sent positions of all commands in the list are calculated,
and the command is executed on the player if the spacetime interval is larger than zero.
Algorithm~\ref{alg:command} illustrates the process.

\begin{algorithm}
\KwInput{\\
  \Indp
  $M_\textrm{current}$ map from player id to current player data\\
  $M_{\textrm{command}}$ map from player id to lists of commands
}

\ForEach{key $i$ in $M_{\textrm{command}}$}{
  Get the integer coordinates $\textbf{S}_i$ of player $i$ from $M_\textrm{current}[i]$\;
  \ForEach{command $C$ in the list $M_{\textrm{command}}[i]$}{
    \If{$\|\textbf{S}_{\textrm{from}} - \textbf{S}_i\| \geq 0$}{
      Update $M_\textrm{current}[i]$ by $f_{\text{target}}$ of the command\;
    }
  }
}
Remove all executed commands\;

\caption{Process command map.}
\label{alg:command}
\end{algorithm}

\subsection{Move players and add afterimages} \label{ssec:move}

Moving players and storing their data requires additional considerations in this simulation framework.
Consider the following example:
\begin{enumerate}
  \item assume player $i$ and player $j$ are located in the same cube,
  \item player $j$ moves to the other cube,
  \item the new information takes time to travel to player $i$, so player $i$ cannot see the new position of player $j$,
  \item player $i$ cannot see the old information of player $j$ either, because player $j$ is no longer there,
  \item player $j$ disappears from the sight of player $i$.
\end{enumerate}

This ``disappearance'' is caused by the problem of the integer-based coordinates used in the computation of player's 3D view.

Consider a more generic situation: suppose player $i$ is located at $\overrightarrow{U_i}$, 
and player $j$ moves from $\overrightarrow{U_j}$ to $\overrightarrow{U_k}$.
Ignoring the possibility of zero time delay, the maximum time player $i$ has to wait to see player $j$
is bounded by Eq.~\ref{eq:deltaT},

\begin{align} ~\label{eq:deltaT}
  \Delta T &= \tau(\overrightarrow{U_i}, \overrightarrow{U_j}) - \tau(\overrightarrow{U_i}, \overrightarrow{U_k}), \\
  &= \left \lceil \frac{|\overrightarrow{U_i} - \overrightarrow{U_j}|}{c} \right \rceil - \left \lceil \frac{|\overrightarrow{U_i} - \overrightarrow{U_k}|}{c} \right \rceil,  \\
  &\leq \left \lceil \frac{|\overrightarrow{U_i} - \overrightarrow{U_j}|}{c} - \frac{|\overrightarrow{U_i} - \overrightarrow{U_k}|}{c} \right \rceil, \\
  &\leq \left \lceil \frac{|\overrightarrow{U_j} - \overrightarrow{U_k}|}{c} \right \rceil, \\
  &=\tau(\overrightarrow{U_j}, \overrightarrow{U_k}).
\end{align}

Therefore, if we include back the possibility where the time delay between player $i$ and player $j$ can be zero,
the maximum duration of the disappearance produced by the movement is bounded by $\Delta T_{\textrm{max}} = \tau((0, 0, 0), (1, 1, 1))$.
To prevent the unrealistic disappearance from happening, the old player data has to stay at the original
position for at least $\Delta T_{\textrm{max}}$ turn, we call this the ``afterimage'' of the player.
Note that afterimages only participate in the 3D view of players, they should not be updated by commands or mechanisms.

Algorithm~\ref{alg:move} does multiple things: it updates the universe time, 
it moves players by their velocities, it synchronizes time of players,
it stores old coordinates to the history of player, it cleans the history if the stored coordinates is too old,
and it adds the current player and afterimages to the latest spatial 3D array in the 4D data array $M_{TXYZ}$.
Since the universe time has been updated, this simulation step has finished, 
the universe should go to the next step and loop over all algorithms in Sec.~\ref{sec:simulation} again.

\begin{algorithm}
\KwInput{\\
  \Indp
  $M_\textrm{current}$ map from player id to current player data\\
  Universe data
}

$T_{\textrm{current}} \gets T_{\textrm{current}} + 1$\;
Initialize a 3D array of maps from player id to lists of player data $M_{XYZ}$\;
\ForEach{data of player $i$ in $M_\textrm{current}$}{
  $t_i \gets T_{\textrm{current}}$\;
  $x_i \gets x_i + v_{ix}$\;
  $y_i \gets y_i + v_{iy}$\;
  $z_i \gets z_i + v_{iz}$\;
  $T_i \gets T_{\textrm{current}}$\;
  $X_i \gets \lfloor x_i \rfloor$\;
  $Y_i \gets \lfloor y_i \rfloor$\;
  $Z_i \gets \lfloor z_i \rfloor$\;
  $g_i \gets g(\overrightarrow{u_i}, \overrightarrow{U_i})$ by Eq.~\ref{eq:group}\;
  \If{coordinates or group is new}{
    Save the previous coordinates to history $H_i$\;
  }
  \ForEach{$(T_i', X_i', Y_i', Z_i')$ in $H_i$}{
    Remove from $H_i$ if $T_{\textrm{current}} - T' > \Delta T_{\textrm{max}}$\;
  }
  Save the new data to $M_{X_i Y_i Z_i}[i]$\;
  \ForEach{$(T_i', X_i', Y_i', Z_i')$ in $H_i$}{
    Find the old player data from $M_{T_i' X_i' Y_i' Z_i'}[i']$\;
    Add the old player data to $M_{X_i' Y_i' Z_i'}[i']$\;
  }
}
Drop the oldest 3D spatial array from $M_{TXYZ}$\;
Add $M_{XYZ}$ as the latest spatial array to $M_{TXYZ}$\;

\caption{Move player and add afterimages.}
\label{alg:move}
\end{algorithm}

\section{Discussion}

The presented algorithms form the backbone of our computational framework, 
``Relativitization''~\cite{relativitization}.
There are technical subtleties that are not discussed here,
such as creating new players, removing dead players,
introducing randomness to models, parallelization of the algorithms, 
generating deterministic outcomes from parallelized simulations with random number generators,
interactive human input to intervene in a simulation, etc.
Nevertheless, the framework implements the major part of the technical subtleties, 
and provides a suitable interface to ease the development of any 4D, relativistic ABM.

It can be interesting to implement a classical ABM into the framework. 
Spatial ABMs with non-local interactions, 
such as the classical flocking model~\cite{reynolds1987flocks},
are particularly suitable.
These models are naturally affected by the time delay
imposed by the speed of light limitation.
Simulating such a model in the Relativitization framework allows us to explore the effects of time delay on the model.

Ultimately, existing ABMs might not be suitable to describe interstellar society.
A solid understanding of social mechanisms and physics,
together with some artistic imagination,
are needed to build inspiring interstellar ABMs.
As a first step,
we have integrated a few social mechanisms to build a big ``model'', which is also a game. 
The ``model'' can be found on the GitHub\footnote{https://github.com/Adriankhl/relativitization} repository of our framework.

Apart from the possibility of implementing different models using the framework, 
the algorithms may also be optimized further.
For example, the iteration in Sec.~\ref{ssec:mechanisms} has a time complexity of $O(mn)$.
A naive alternative implementation to iterate over all the combinations of players could change the complexity to $O(n^2)$,
which could have better performance when the density of players is low.
We leave these potential improvements to future research.

\section{Conclusion}

In this paper, we have presented a set of algorithms to implement ABM simulations in a 4D, relativistic spacetime.
Based on these algorithms, we have developed a simulation framework we call ``Relativitization''~\cite{relativitization}.
Our framework will lower the barrier of entry for social scientists
to apply their expertise to explore the interstellar future of human civilization.
We hope our framework can be used to initiate meaningful and academically interesting discussions
about our future.

\section*{Acknowledgement}
We thank Diego Garlaschelli, Alexandru Babeanu, Michael Szell, and all QSS members of the CWTS institute for useful discussion. 
%
%
\printbibliography

@software{relativitization,
  author       = {Lai, Kwun Hang},
  title        = {Relativitization},
  year         = 2022,
  doi          = {10.5281/zenodo.6120765},
  url          = {https://doi.org/10.5281/zenodo.6120765},
  howpublished = {https://github.com/Adriankhl/relativitization},
}

@software{netlogo,
  author = {Wilensky, U.},
  title = {NetLogo},
  howpublished = {http://ccl.northwestern.edu/netlogo/},
  year={1999},
  institute = {Center for Connected Learning and Computer-Based Modeling, Northwestern University, Evanston, I},
}

@InProceedings{python-mesa-2020,
  author={Kazil, Jackie and Masad, David and Crooks, Andrew},
  editor={Thomson, Robert and Bisgin, Halil and Dancy, Christopher and Hyder, Ayaz and Hussain, Muhammad},
  title={Utilizing Python for Agent-Based Modeling: The Mesa Framework},
  booktitle={Social, Cultural, and Behavioral Modeling},
  year={2020},
  publisher={Springer International Publishing},
  address={Cham},
  pages={308--317},
  isbn={978-3-030-61255-9}
}

@article{Agents2021,
  author = {George Datseris and Ali R. Vahdati and Timothy C. DuBois},
  title ={Agents.jl: a performant and feature-full agent-based modeling software of minimal code complexity},
  journal = {SIMULATION},
  volume = {0},
  number = {0},
  pages = {00375497211068820},
  year = {2021},
  doi = {10.1177/00375497211068820},
  URL = {https://doi.org/10.1177/00375497211068820},
  eprint = {https://doi.org/10.1177/00375497211068820},
}

@article{gray2015fermi,
  title={The Fermi paradox is neither Fermi's nor a paradox},
  author={Gray, Robert H},
  journal={Astrobiology},
  volume={15},
  number={3},
  pages={195--199},
  year={2015},
  publisher={Mary Ann Liebert, Inc. 140 Huguenot Street, 3rd Floor New Rochelle, NY 10801 USA}
}

@book{edmonds2015simulating,
  author = {Edmonds, Bruce and Meyer, Ruth},
  publisher = {Springer},
  title = {Simulating social complexity},
  year = {2015}
}

@article{wright2020dyson,
  title={Dyson spheres},
  author={Wright, Jason T},
  journal={arXiv preprint arXiv:2006.16734},
  year={2020}
}

@article{gray2020extended,
  title={The Extended Kardashev Scale},
  author={Gray, Robert H},
  journal={The Astronomical Journal},
  volume={159},
  number={5},
  pages={228},
  year={2020},
  publisher={IOP Publishing}
}

@article{pal2020review,
  title={A review of platforms for the development of agent systems},
  author={Pal, Constantin-Valentin and Leon, Florin and Paprzycki, Marcin and Ganzha, Maria},
  journal={arXiv preprint arXiv:2007.08961},
  year={2020}
}

@inproceedings{reynolds1987flocks,
  title={Flocks, herds and schools: A distributed behavioral model},
  author={Reynolds, Craig W},
  booktitle={Proceedings of the 14th annual conference on Computer graphics and interactive techniques},
  pages={25--34},
  year={1987}
}

\end{document}